# Controlling of the Dirac band states of Pb-deposited graphene by using work function difference

Y. Tsujikawa[1], M. Sakamoto[1], Y. Yokoi[1], M. Imamura[2], K. Takahashi[2], R. Hobara[3], T. Uchihashi[4], and A. Takayama[1]*

[1]*Department of Physics and Applied Physics, Waseda University, Shinjuku, Tokyo, 169-8555, Japan*

[2]*Synchrotron Light Application Center, Saga University, Tosu, Saga, 841-0005, Japan*

[3]*Department of Physics, University of Tokyo, Bunkyo, Tokyo 113-0033, Japan*

[4] *International Center for Materials Nanoarchitectonics (WPI-MANA), National Institute for Materials Science, Tsukuba, Ibaraki, 305-0044, Japan*

CORRESPONDING AUTHOR

Name: Akari Takayama

Address: Department of Physics and Applied of Physics, Waseda University,

3-4-1 Okubo, Shinjuku-ku, Tokyo, 169-8555, Japan

TEL: +81-3-5286-2981

E-mail address: a.takayama@waseda.jp

**Abstract**

We have performed scanning tunneling microscope (STM) and angle-resolved photoemission spectroscopy (ARPES) in Pb-deposited bilayer Graphene (BLG) on SiC(0001) substrate to investigate the dependence of the electronic structures on Pb-deposition amount. We have observed that the Pb atoms form islands by STM and the π bands of the BLG shift toward the Fermi level by ARPES. This hole-doping-like energy shift is enhanced as the amount of Pb is increased, and we were able to tune the Dirac gap to the Fermi level by 4 ML deposition. Considering the band dispersion, we suggest that hole-doping-like effect is related to the difference between the work functions of Pb islands and BLG/SiC; the work function of BLG/SiC is lower than that of Pb. Our results propose an easy way of band tuning for graphene with appropriate selection of both the substrate and deposited material.

Graphene has been of interest for its remarkable properties such like massless charge carriers [1,2] and quantum hall effects [3] caused by unique electron structures. One of the challenges in developing graphene devices is the controlling of band gap. For this purpose, many studies [4-8] have been performed. It is well known that difference of potential between top surface and bottom surface makes a gap; namely, substrate effect is very important to open a gap. The electronic states of the Dirac cone with gap have been studied successfully on SiC [6,7] and metal substrates [8]. Meanwhile the modulation of potential by another materials grown on the graphene is also an important issue. Generally, graphene grown on SiC is metallic with a partially occupied $\pi*$ band below the Fermi level ($E_F$) because of the charge (electron) transfer from the SiC substrate and the buffer layer to the graphene sheet [6,7]. In contrast a semiconducting character of graphene is required when we utilize a graphene grown on SiC for application. The semiconducting nature of graphene grown on SiC is recovered by the hydrogen-passivation of dangling bonds at the buffer layer [9]. A hole-doping-like effect is also caused by intercalation such as Fe, Si and so on [10-12].

Yutsever *et al.* [12] reported a result of Pb-intercalated monolayer graphene on SiC(0001) with annealing, and they suggest that a hole-doping-like effect is caused by decoupling of the buffer layer from the substrate by Pb-termination. On the other hand, several studies show results that Pb atoms form islands on graphene because of their weak interaction [13-16]. Yet direct experimental studies conforming the effect of Pb islands on graphene are still few, while there are experiments for intercalating Pb between layers [12,17]. It is unclear that Pb atoms without intercalation provide either electrons or holes to graphene. Understanding and controlling the band structure of Pb-deposited graphene will give a new research field in developing graphene device.

In this study we have carried out scanning tunneling microscope (STM) and angle-resolved photoemission spectroscopy (ARPES) measurement to investigate the surface structure and band dispersion of Pb grown on bilayer graphene (BLG) on SiC. We found that Pb-deposited BLG show a hole-doping-like behavior, tuning the chemical potential to be located in the Dirac gap for 4 ML Pb-deposition. We suggest that this hole-doping-like behavior is caused by the difference between the work functions of Pb and BLG/SiC. This will be an efficient way of developing graphene devices considering its easiness of just depositing Pb on graphene/SiC.

The BLG samples were prepared on an n-type Si-rich 6H-SiC(0001) substrate (dopant density $1\times10^{18}$-$1\times10^{19}$ /cm$^3$) in an ultrahigh vacuum (UHV) chamber. First the SiC substrate was degassed for more than 6 hours, then heated up to 1300°C for 5 minutes and 1500°C for additional 5 minutes. By precisely controlling the heating temperature and the duration time, we fabricated a bilayer graphene [18]. Pb deposition on the BLG was conducted in the UHV chamber by molecular beam epitaxy method. The deposition rate of Pb was calibrated by a pre-experiment on Si, and the Pb coverage was controlled by the deposition time at a constant deposition rate. All samples were checked by low-energy electron diffraction (LEED) measurement. We confirmed that the LEED pattern of graphene still remained after Pb deposition. The STM images were acquired in the constant current mode with an electrochemically etched W tip at room temperature. The Nanotec Electronica WSxM software was used to process the STM images [19]. The ARPES measurements were performed at BL13 in the Saga Light Source (SAGA-LS) using synchrotron radiation and hemispherical electron-energy analyzer [20]. We carried out the ARPES measurements under ultrahigh vacuum (~$1\times10^{-8}$ Pa). The sample temperature and the photon energy were set $T = 100$ K and $h\nu$

= 40 eV, respectively.

We first report on results of STM measurement. Figure 1(a) shows the STM image of 4ML-Pb/graphene/SiC. We found that Pb atoms formed small islands on the terrace of graphene instead of films without wetting layer. It indicates that Pb islands growth on graphene by Volmer-Weber growth mode. The shape and size of these Pb islands are close to previous studies of Pb-graphene [13,14]. The most common size and height of Pb islands are ~2000 nm$^2$ and 3 nm (~10 ML), respectively. We confirmed that the total deposition amount which is calculated from the density and the height of Pb islands of the STM image is consistent with estimated deposition amount. We also confirmed that the density and heights of Pb islands increased as the amount of deposition increased above 4 ML and there is no wetting layer. This growth process of Pb is also identical to previous studies [13,14].

To verify the change in electron states before and after Pb deposition, we measured band dispersions by ARPES. Figures 2(a)-(d) are the dependence of the characteristic Dirac bands on the deposition amount around the K point for 0, 1.0, 1.5 and 4.0 ML-Pb on BLG, respectively. One can clearly see that pristine BLG has two π bands and one π∗ band as reported in previous studies [6]. By further increasing the Pb amount, we found that the whole bands shift toward the Fermi level. This means that Pb atoms work as hole dopants. We also observed that the Dirac band became broader by Pb deposition, suggesting that Pb atoms may introduce defects into the graphene during deposition. Even so, for 4ML-Pb/graphene, the gap of the Dirac cone is still opened, and the $E_F$ is located within the gap. Considering the size of each islands (~100 nm) being much smaller than that of the spot diameter of ARPES (~0.05 mm), it should be noted that the band information is collected at the macroscopic scale. If electron state is different between pristine BLG and BLG under Pb island, we must observe two kinds of the Dirac bands individually; one is original Dirac band from pristine BLG without Pb, the other is hole-doped Dirac band from BLG under Pb island. However, we measured just one kind of the Dirac band. We have also measured position dependence in the band structure and observed uniform band structure. The fact indicates that hole is doped in whole graphene uniformly. To discuss the amount of this energy shift quantitatively, we compared each energy distribution curves (EDCs) at the K point as shown in Fig. 2(e). We analyzed the experimental results to determine the states above $E_F$ by utilizing the finite distribution of the Fermi-Dirac distribution (FD) function. The Dirac point (DP), which is defined as the midpoint of the top of the $\pi$ band and the bottom of the $\pi*$ band, is used to clearly define the energy shift (see blue triangles in Fig. 2(e) to indicate the energy of the DP). For pristine BLG, the EDC mainly consists of three peaks; the bottom of $\pi*$ band at 0.22 eV, the top of outer $\pi$ band at 0.39 eV, and the inner of $\pi$ band at 0.72 eV, respectively. The peak of $\pi*$ band disappears for 4 ML-Pb/graphene, while the $\pi$ and $\pi*$ band are still observed for 1 ML-Pb/graphene.

      Figures 3(a)-(d) also show the dependence of the valence band on the deposition amount around the $\Gamma$ point for 0, 1.0, 1.5 and 4.0 ML-Pb on BLG, respectively. Now we focus on the band structure at the $\Gamma$ point. Although each band dispersion appears the same at first sight, with a careful look at the EDCs in Fig. 3(e), we were also able to observe the $\pi$ bands sifting to the Fermi Energy at the $\Gamma$ point as the Pb deposition increases (see black triangles to indicate the peak position). The specific amount of energy shift at the K and the $\Gamma$ point estimated from EDCs are summarized in Table1. Intriguingly, the absolute values of energy shifts of $\pi$ bands at the $\Gamma$ point is different from that at the K point, *i.e.*, the magnitude of energy shift at the

Γ point is larger than that of the K point. Our result indicates that the width of the π band between the bottom (the Γ point) and top (the K point) vary depending on the deposition amount of Pb, resulting in a squeezing of the π band by Pb evaporations. We also comment that the σ bands also shifted due to Pb depositions although this change was very small compared to that of π bands. The π band is constructed by $p_z$ orbital components, while the σ band is constructed by the $sp_2$ hybrid orbital components [21]. Therefore, it is considered that the π band is affected by the out-of-plane potential gradient ($E_z$) more strongly than the σ band, while the $E_z$ changes with Pb deposition.

We discuss the details of changes in the band structure caused by Pb deposition on graphene. Our results show that Pb atoms affect graphene as hole-dopants, although Pb is metal which gives electrons in usual situations. Similar hole-doping-like energy shifts were reported by Yutsever *et al.* [12] as mentioned above. On the other hand, comparing their result with our data, we did not observe any additional bands formed by Pb-termination. Our results imply that Pb atoms are not directly intercalated if no additional annealing is taken after Pb deposition. Hence, we present a different explanation by simply modeling the interaction between SiC-graphene and Pb in terms of work function [22-24]. Figure 4 illustrates the schematic diagram for potential of BLG/SiC and Pb. In the case of free-standing BLG, the Dirac point is located at the $E_F$ with no gaps. As seen in Fig. 2(a), the Dirac point of the BLG on SiC is located 0.33 eV below $E_F$ and a gap exist. As mentioned above, this is attributed to the effect of the SiC substrate [6]. The Dirac point from the Vacuum level can be said to be around 4.5eV in regardless of the dopant volume of the SiC substrate [25-27], although the work function of graphene also changes sensitively by the effect of the substrate [25-28]. Considering from the Dirac point of our pristine graphene, we estimated the work function of our BLG to be 4.17 eV. This is in good consistent with the report from Gugel et al. that says the work function of BLG on high-doped SiC to be 4.19 eV. This value of work function is lower than of Pb (4.25 eV) [29]. When graphene interacts with Pb and SiC, charge transfer among the three comes to equilibrium and these Fermi levels are aligned. Since the density of state (DOS) of graphene around the $E_F$ is much smaller than that of Pb, this equilibrium is achieved by lowering the fermi level of graphene. As a result, the band dispersion varies as if holes are doped to BLG. Thus, we conclude that hole-doping-like energy shift by Pb deposition occurs without intercalation. This charge transfer between graphene and Pb yeilds a strong out-of-plane electric field ($E_z$), thus it is compatible with the result in the squeezing of the π band. This is consistent with a previous study that predicts an electron doping for the Pb islands on graphene [13] since the graphene was assumed to be free-standing having a higher work function than the Pb. In future, high-energy resolution experiments are required for quantitative analysis to investigate different hole-doping-like behaviors of intercalated-Pb and Pb islands.

In summary, we have performed STM and ARPES measurement of Pb-deposited BLG on SiC substrate. We observed hole-doping-like band shifts and a squeezing of the π band. We conclude that the hole-doping-like behavior is caused by the difference between the work functions of Pb and BLG/SiC. Our results suggest the importance of the work function difference regarding both the substrate and deposited material in controlling electron state of the Dirac cone. From this point of view, Ag, Cu and Pt also have been considered as the promising candidate as hole dopants because of their work function [23]. We emphasize that Pb is one of the best materials for BLG/SiC because the $E_F$ can be tuned within the gap due to its suitable work function.

**Acknowledgements**

We thank S. Hasegawa (University of Tokyo) for his participation on discussion. This work was supported by JSPS (Grant No.19H04398). A. T acknowledges the financial support by the NIMS Joint Research Hub Program. The photoemission experiments were performed at Saga University Beamline (SAGA-LS/BL13) with a proposal of H30-204P and H30-307P under the support of the Ministry of Education, Culture, Sports, Science and Technology (MEXT), Japan.

24. A. Varykhalov. M. R. Scholz, T. K. Kim, and O. Rader, Phys. Rev. B 82, 121101 (2010).

25. D. Guel, D. Niesner, C. Eickhoff, S. Wagner, M. Weinelt, and T. Fauster, 2D mater. 2, 045001, (2015).

26. S. Mammadov, J. Ristein, J. Krone, C. Raidel, M. Wanke, V. Wiesmann, F. Speck, and T. Seyller, 2D Mater. 4, 015043 (2017).

27. O. Renault, A. M. Pascon, H. Rotella, K. Kaja, C. Mathieu, J. E. Rault, P. Blaise, T. Poiroux, N. Barrett, and L. R. C. Fonseca, J. Phys. D: Appl. Phys. 47, 295303 (2014).

28. C. Mathieu, N. Barrett, J. Rault, Y. Y. Mi, B. Zhang, W. A. de Heer, C. Berger, E. H. Conrad, and O. Renault, Phys. Rev. B 83, 235436 (2011).

29. H. B. Michaelson, J. Appl. Phys. 48, 4729 (1977).
12

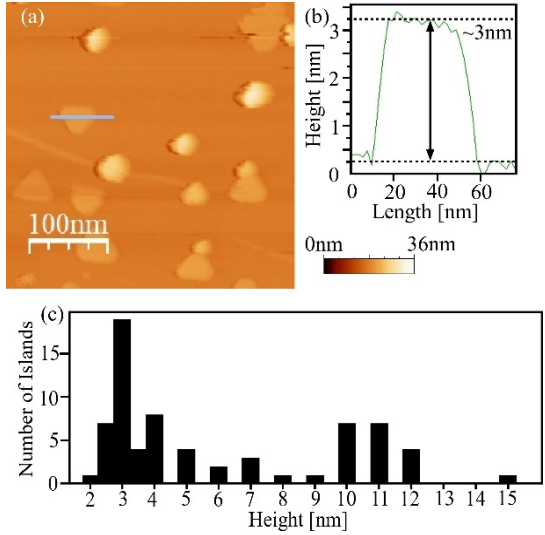

**Fig. 1.** (a) STM images of Pb islands constructed on graphene for 4 ML of Pb deposited. ($U$ = 1.5 V, $I$ = 80 pA). (b) Height Profile along the blue line of (a). (c) Rough distribution of each islands heights in same sample of (a).

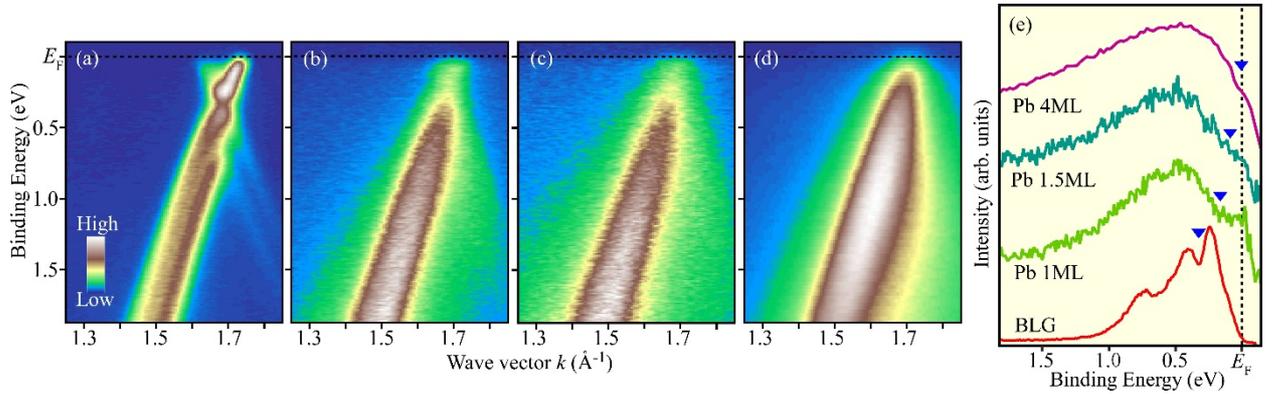

**Fig. 2.** (a)-(d) Pb-deposition amount dependence of band dispersion around the K point of graphene for (a) pristine BLG (0 ML), (b) 1.0 ML, (c) 1.5 ML, and (d) 4.0 ML, respectively. (e) EDC of Graphene and EDCs divided by the FD function at the K point for (a)-(d). The blue triangles indicate the DP in Table1.

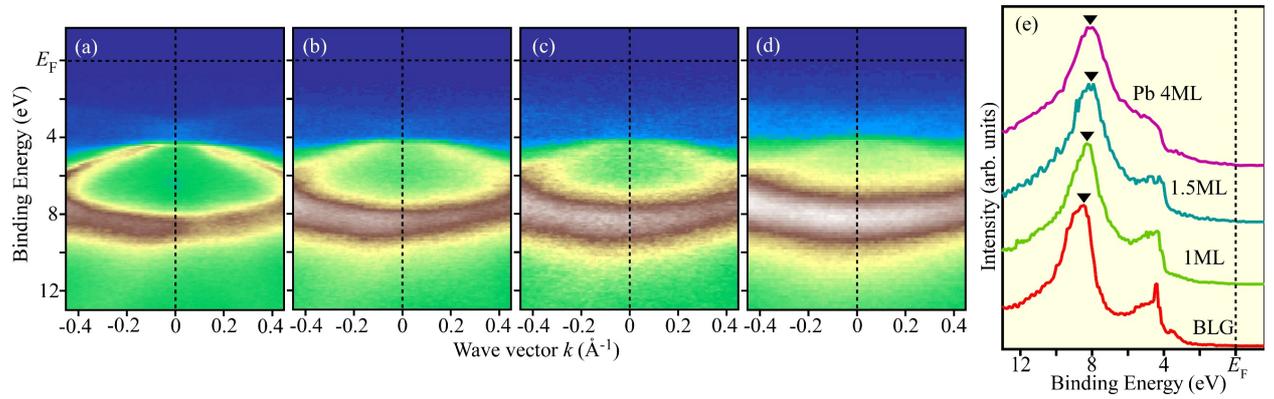

**Fig. 3.** (a)-(d) Pb-deposition amount dependence of band dispersion around the Γ point of graphene for (a) pristine BLG (0 ML), (b) 1.0 ML, (c) 1.5 ML, and (d) 4.0 ML, respectively. (e) EDCs at the Γ point for (a)-(d). The left peak represents the π band and the right band represents the σ band. Black triangles indicate the bottom of the π band at the Γ point.

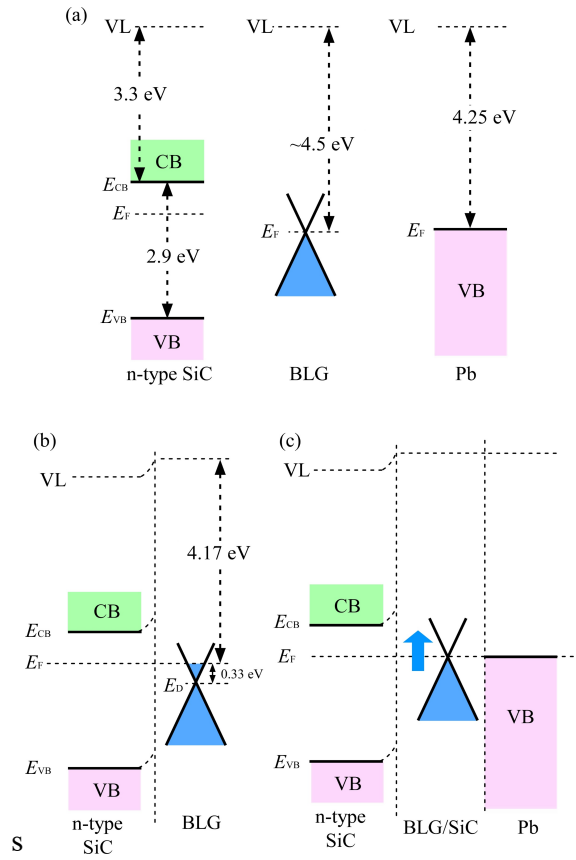

**Fig. 4.** Schematic diagram of energy band of Pb-deposited BLG on SiC(0001) substrate. (a) the normal potential of SiC, BLG and Pb before they contact. (b), (c) A potential image of BLG on SiC and Pb-deposited BLG on SiC (after they contact).

Table 1. Energy shift of DP at the K and the bottom of the π band at the Γ point.

|  | K (eV) [DP] | Γ (eV) [Bottom of π band] |
|---|---|---|
| **BLG** | 0.33 | 8.46 |
| **1 ML-Pb/BLG** | 0.17 | 8.27 |
| **1.5 ML-Pb/BLG** | 0.09 | 8.09 |
| **4 ML-Pb/BLG** | ~0 | 7.97 |